\documentclass{article}

 \usepackage[preprint]{neurips_2025}

\usepackage[utf8]{inputenc} 
\usepackage[T1]{fontenc}    
\usepackage{hyperref}       
\usepackage{url}            
\usepackage{booktabs}       
\usepackage{amsfonts}       
\usepackage{nicefrac}       
\usepackage{microtype}      
\usepackage{xcolor}         
\usepackage{setspace}

\renewcommand{\em}{\it}

\newcommand{\ignore}[1]{}

\def\cfigure[#1,#2,#3]{
\begin{figure}
\vspace*{0mm}
\begin{center}

\includegraphics[width=3in]{#1} 
 
\vspace*{-3mm}\caption[]{#2
} \label{#3}
 
\vspace*{-5mm}
\end{center}
\end{figure}}

\def\cfigurefour[#1,#2,#3]{
\begin{figure}
\vspace*{0mm}
\begin{center}

\includegraphics[width=4in]{#1} 
 
\vspace*{-3mm}\caption[]{#2
} \label{#3}
 
\vspace*{-5mm}
\end{center}
\end{figure}}

\def\cfiguretemp[#1,#2,#3]{
\begin{figure}
\vspace*{0mm}
\begin{center}

\includegraphics[width=3.5in]{#1} 
 
\vspace*{-3mm}\caption[]{#2
} \label{#3}
 
\vspace*{-5mm}
\end{center}
\vspace*{-2mm}
\end{figure}}

\def\wfigure[#1,#2,#3]{
\begin{figure*}
\vspace*{0mm}
\begin{center}
 \includegraphics[width=\textwidth]{#1} 
 \vspace*{-3mm}\caption[]{#2
} \label{#3}
 
\end{center}
\end{figure*}}

\def\threefigure[#1,#2,#3,#4,#5]{
\begin{figure*}
\vspace*{0mm}
\begin{center}

\begin{tabular}{ccc}
\includegraphics[width=2in]{#1} & \includegraphics[width=2in]{#2} &  \includegraphics[width=2in]{#3} \\
(a) & (b) & (c) \\
\end{tabular}

\vspace*{-3mm}\caption[]{#4
} \label{#5}

\vspace*{-5mm}
\end{center}
\vspace*{-2mm}
\end{figure*}}

\def\dcfigure[#1,#2,#3,#4,#5,#6]{
{
\begin{figure*}
\begin{center}
\begin{minipage}[c]{\columnwidth}{
\includegraphics[width=\columnwidth]{#1} 
\vspace*{0mm}\caption[]{#2} \label{#3} \
}\end{minipage}\hspace*{\columnsep}\
\begin{minipage}[c]{\columnwidth}{
\includegraphics[width=\columnwidth]{#4} 
\vspace*{0mm}\caption[]{#5}\label{#6} \
}\end{minipage}
\end{center}
\end{figure*}
}
}

\def\tableByTable[#1,#2,#3,#4,#5,#6]{
{
\begin{table*}
\begin{center}
\begin{minipage}[c]{3in}{
\centering
{#1}
\vspace*{0mm}\tabcaption[]{#2}\label{#3} \
}\end{minipage}\hspace*{\columnsep}\
\begin{minipage}[c]{3in}{
\centering
{#4}
\vspace*{0mm}\tabcaption[]{#5}\label{#6} \
}\end{minipage}
\end{center}
\end{table*}
}
}

\def\figureByTable[#1,#2,#3,#4,#5,#6]{
{
\begin{figure*}
\begin{center}
\begin{minipage}[c]{3in}{
\centering
\includegraphics[width=\textwidth]{#1}
\vspace*{0mm}\figcaption[]{#2} \label{#3} \
}\end{minipage}\hspace*{\columnsep}\
\begin{minipage}[c]{3.3in}{
\centering
{#4}
\vspace*{0mm}\tabcaption[]{#5}\label{#6} \
}\end{minipage}
\end{center}
\end{figure*}
}
}

\def\tableByFigure[#1,#2,#3,#4,#5,#6]{
{
\begin{figure*}
\begin{center}
\begin{minipage}[c]{4.3in}{
\centering
{#1}
\vspace*{0mm}\tabcaption[]{#2} \label{#3} \
}\end{minipage}\hspace*{\columnsep}\
\begin{minipage}[c]{2.2in}{
\centering
\includegraphics[width=\textwidth]{#4}
\vspace*{-0.35in}\caption[]{#5}\label{#6} \
}\end{minipage}
\end{center}
\end{figure*}
}
}

\def\doublecfigure[#1,#2,#3,#4]{
{
\begin{figure}
\begin{center}
\begin{minipage}[c]{1.5in}{
\begin{center}
\includegraphics[width=1.5in]{#1}
\end{center}
}\end{minipage}\hspace*{1em}\
\begin{minipage}[c]{1.5in}{
\begin{center}
\includegraphics[width=1.5in]{#2}
\end{center}
}\end{minipage}
\vspace*{0mm}\caption[]{#3} \label{#4} \
\end{center}
\end{figure}
}
}

\def\qcfigure[#1,#2,#3,#4,#5,#6]{
{
\begin{figure*}
\vspace*{0.2in}\
\begin{center}
\begin{minipage}[c]{3in}{
\includegraphics[width=3in]{#1} 
\vspace*{-3mm}
}
\end{minipage}\hspace*{0.5in}\
\begin{minipage}[c]{3in}{
\includegraphics[width=3in]{#2} 
\vspace*{-3mm}
}\end{minipage}

\begin{minipage}[c]{3in}{
\includegraphics[width=3in]{#3} 
\vspace*{-3mm}
}
\end{minipage}\hspace*{0.5in}\
\begin{minipage}[c]{3in}{
\includegraphics[width=3in]{#4} 
\vspace*{-3mm}
}\end{minipage}
\end{center}
\caption[]{#5}\label{#6}
\end{figure*}
}
}

\def\twfigure[#1,#2,#3,#4,#5]{
{
\begin{figure*}
\vspace*{0.2in}\
\begin{center}
\begin{minipage}[c]{6.5in}{
\includegraphics[width=6.5in]{#1} 
\vspace*{-3mm}
}
\end{minipage}

\begin{minipage}[c]{6.5in}{
\includegraphics[width=6.5in]{#2} 
\vspace*{-3mm}
}\end{minipage}

\begin{minipage}[c]{6.5in}{
\includegraphics[width=6.5in]{#3} 
\vspace*{-3mm}
}
\end{minipage}
\end{center}
\caption[]{#4}\label{#5}
\end{figure*}
}
}

\def\dwfigure[#1,#2,#3,#4]{
{
\begin{figure*}
\vspace*{0.2in}\
\begin{center}
\begin{minipage}[c]{6.5in}{
\includegraphics[width=6.5in]{#1} 
\vspace*{-3mm}
}
\end{minipage}

\begin{minipage}[c]{6.5in}{
\includegraphics[width=6.5in]{#2} 
\vspace*{-3mm}
}\end{minipage}

\end{center}
\caption[]{#3}\label{#4}
\end{figure*}
}
}

\def\dssfigure[#1,#2,#3,#4,#5,#6]{
{
\begin{figure*}
\vspace*{0.2in}\
\begin{center}
\begin{minipage}[c]{4in}{
\includegraphics[width=4in]{#1}
\vspace*{-3mm}\caption[]{#2} \label{#3} \
}\end{minipage}\hspace*{0.5in}\
\begin{minipage}[c]{2in}{
\includegraphics[width=2in]{#4}
\vspace*{-3mm}\caption[]{#5}\label{#6} \
}\end{minipage}
\end{center}
\vspace*{-0.4in}\
\end{figure*}
}
}

\def\dsfigure[#1,#2,#3,#4,#5,#6]{
{
\begin{figure*}
\vspace*{0.2in}\
\begin{center}
\begin{minipage}[c]{3in}{
\includegraphics[width=3in]{#1}
\vspace*{-3mm}\caption[]{#2} \label{#3} \
}\end{minipage}\hspace*{0.5in}\
\begin{minipage}[c]{3in}{
\hspace*{0.5in}\
\includegraphics[height=3in]{#4}
\vspace*{-3mm}\caption[]{#5}\label{#6} \
}\end{minipage}
\end{center}
\vspace*{-0.4in}\
\end{figure*}
}
}

\def\dsyfigure[#1,#2,#3,#4,#5,#6]{
{
\begin{figure*}
\vspace*{0.2in}\
\begin{center}
\begin{minipage}[c]{2.5in}{
\includegraphics[height=2.5in]{#1}
\vspace*{-3mm}\caption[]{#2} \label{#3} \
}\end{minipage}\hspace*{0.5in}\
\begin{minipage}[c]{2.5in}{
\includegraphics[height=2.5in]{#4}
\vspace*{-3mm}\caption[]{#5}\label{#6} \
}\end{minipage}
\end{center}
\vspace*{-0.4in}\
\end{figure*}
}
}

\def\dyfigure[#1,#2,#3,#4,#5,#6]{
{
\begin{figure*}
\vspace*{0.2in}\
\begin{center}
\begin{minipage}[c]{3in}{
\includegraphics[height=3in]{#1} 
\vspace*{-3mm}\caption[]{#2} \label{#3} \
}\end{minipage}\hspace*{0.5in}\
\begin{minipage}[c]{3in}{
\includegraphics[height=3in]{#4} 
\vspace*{-3mm}\caption[]{#5}\label{#6} \
}\end{minipage}
\end{center}
\vspace*{-0.4in}\
\end{figure*}
}
}

\def\dyoldfigure[#1,#2,#3,#4,#5,#6]{
{
\begin{figure*}
\vspace*{0.2in}\
\begin{center}
\begin{minipage}[c]{3in}{
\epsfysize=2.0in\
\hspace{0.5in}\
\epsfbox{#1}
\vspace*{-3mm}\caption[]{#2} \label{#3} \
}\end{minipage}\hspace*{0.25in}\
\begin{minipage}[c]{3in}{
\epsfysize=2.0in\
\hspace{0.5in}\
\epsfbox{#4}
\vspace*{-3mm}\caption[]{#5}\label{#6} \
}\end{minipage}
\end{center}
\vspace*{-0.4in}\
\end{figure*}
}
}

\def\cfiguredouble[#1,#2,#3,#4]{
\begin{figure}
\vspace*{0.2in}\
\begin{center}
\begin{minipage}[c]{1.5in}{
\epsfxsize=1.5in\
\epsfbox{#1}
}\end{minipage}\hspace*{0.1in}\
\begin{minipage}[c]{1.5in}{
\epsfxsize=1.5in\
\vspace{0.1in}\epsfbox{#2}
}\end{minipage}\vspace*{-0.10in} \caption[]{#3}\label{#4}
\end{center}
\vspace*{-0.4in}\
\end{figure}
}

\def\wpfigure[#1,#2,#3,#4]{
\begin{figure*}
\vspace*{4mm}
\begin{center}

\includegraphics[width=#4]{#1} 

\vspace*{-3mm}\caption[]{#2
} \label{#3}

\vspace*{-5mm}
\end{center}
\end{figure*}}

\def\wprfigure[#1,#2,#3,#4,#5]{
\begin{figure*}
\vspace*{4mm}
\begin{center}

\includegraphics[width=#4, angle=#5]{#1} 

\vspace*{-3mm}\caption[]{#2
} \label{#3}

\vspace*{-5mm}
\end{center}
\end{figure*}}

\def\DoubleFigureWSlide[#1,#2,#3,#4,#5,#6,#7,#8,#9]{
\begin{figure*}
\vspace*{#9}
\begin{center}
\begin{minipage}{#4}
\includegraphics[width=#4]{#1}
\vspace*{-3mm}\caption{#2
}\label{#3}
\end{minipage}
\hspace{2em}
\begin{minipage}{#8}
\includegraphics[width=#8]{#5}
\vspace*{-3mm}\caption{#6
}\label{#7}
\end{minipage}
\vspace*{-5mm}
\end{center}
\end{figure*}
}

\def\DoubleFigureW[#1,#2,#3,#4,#5,#6,#7,#8]{
\begin{figure*}
\vspace*{0in}
\begin{center}
\begin{minipage}{#4}
\includegraphics[width=#4]{#1}
\vspace*{-3mm}\caption{#2
}\label{#3}
\end{minipage}
\hspace{2em}
\begin{minipage}{#8}
\includegraphics[width=#8]{#5}
\vspace*{-3mm}\caption{#6
}\label{#7}
\end{minipage}
\vspace*{-5mm}
\end{center}
\end{figure*}
}

\def\DoubleFigureWHack[#1,#2,#3,#4,#5,#6,#7,#8]{
\begin{figure*}
\vspace*{0in}
\begin{center}
\begin{minipage}{3in}
\includegraphics[width=#4]{#1}
\vspace*{-3mm}\caption{#2
}\label{#3}
\end{minipage}
\hspace{2em}
\begin{minipage}{3in}
\includegraphics[width=#8]{#5}
\vspace*{-3mm}\caption{#6
}\label{#7}
\end{minipage}
\vspace*{-5mm}
\end{center}
\end{figure*}
}

\def\ddcfigure[#1,#2,#3,#4]{
\begin{figure*}
\vspace*{0.2in}\
\begin{center}
\begin{minipage}[c]{\columnwidth}{
\includegraphics[width=\columnwidth]{#1} 
}\end{minipage}\hspace{0.5in}\
\begin{minipage}[c]{\columnwidth}{
\includegraphics[width=\columnwidth]{#2} 
}\end{minipage} \caption[]{#3}\label{#4}
\end{center}
\end{figure*}
}

\def\ddcfigureSlide[#1,#2,#3,#4,#5]{
\begin{figure*}
\vspace*{#5}\
\begin{center}
\begin{minipage}[c]{3in}{
\includegraphics[height=3in]{#1} 
}\end{minipage}\hspace{0.5in}\
\begin{minipage}[c]{3in}{
\includegraphics[height=3in]{#2} 
}\end{minipage}\vspace*{-0.10in} \caption[]{#3}\label{#4}
\end{center}
\vspace*{-0.4in}\
\end{figure*}
}

\def\cxfigure[#1,#2,#3]{
\begin{figure}
\vspace*{4mm}
\begin{center}
 
\epsfxsize=2.5in\
\epsfbox{#1}\
 
\vspace*{-0.10in}\caption[]{#2
} \label{#3}
 
\vspace*{-5mm}
\end{center}
\vspace*{-2mm}
\end{figure}}

\newcommand{\beforecaption}{\vspace{-.15cm}\begin{spacing}{0.85}}
\newcommand{\aftercaption}{\vspace{-.45cm}\end{spacing}}
\newcommand{\mycaption}[3]{\beforecaption\caption{\label{#1}{\bf #2} \em\small #3}\aftercaption}

\newcommand{\eg}{\textit{e.g.}}
\newcommand{\ie}{\textit{i.e.}}

\newcommand{\KB}{\,KB}

\newcommand{\mus}{\mbox{$\mu s$}}

\newcommand{\boldunderpara}[1]{\noindent{\underline{\textbf{#1}}}}

\newcommand{\sys}{Memix}

\usepackage{tikz}

\title{An Early Exploration of Deep-Learning-Driven Prefetching for Far Memory}

\author{%
  Yutong Huang$^{1}$ \quad
  Zhiyuan Guo$^{1}$ \quad
  Yiying Zhang$^{1,2}$ \\
  $^{1}$University of California, San Diego \quad
  San Diego, CA, USA \\
  $^{2}$GenseeAI Inc. \quad San Diego, CA, USA \\
  \texttt{\{yutonghuang,z9guo,yiying\}@ucsd.edu}
}

\begin{document}

\maketitle

\begin{abstract}
Far-memory systems, where applications store less-active data in more energy-efficient memory media, are increasingly adopted by datacenters.
However, applications are bottlenecked by on-demand data fetching from far- to local-memory.
We present \textbf{\textit{\sys}},
a far-memory system that embodies a deep learning–system co-design for efficient and accurate prefetching, minimizing on-demand far-memory accesses.
One key observation is that memory accesses are shaped by both application semantics and runtime context, providing an opportunity to optimize each independently.
Preliminary evaluation of \sys\ on data-intensive workloads shows that it outperforms the state-of-the-art far-memory system by up to 42\%. 
\end{abstract}
\section{Introduction}

Applications in today’s datacenters,
such as data analytics, large-language model inference, and vector information retrieval~\cite{chen2016xgboost,han2024graph,vllm_pageattention},
exhibit unprecedented memory demands.
However, scaling server-local memory (CPU DRAM, GPU HBM) to meet these demands is untenable due to limited capacity, high cost per GB, and steep energy overheads.
In response, major datacenter providers such as Google and Microsoft increasingly augment hosts with more energy-efficient, slower, and non-local memory tiers, \eg, network-attached pooled memory and non-volatile memory~\cite{chen2023pond,dex2024vldb}. Such memory systems are commonly referred to as {\em far memory}.
Far-memory systems offer lower cost/GB, higher energy efficiency, lower carbon footprint, and elastic capacity.
Prior works show that the reuse of out-dated DDR4 memory as far memory could save the carbon footprint of the datacenter up to 20\% \cite{osdi24_cxlvm,lyu23hotcarbon}. 
The main challenge of practicing far memory is the large latency gap: An RDMA access to remote memory can be over 20× slower than local memory~\cite{rdma_latency}, resulting in substantial slowdowns and negating energy-saving benefits~\cite{chen2023pond,fastswap,hermit}.

%

An effective way to hide this latency is prefetching, which predicts future accesses and fetches data from far to local memory in advance.
However, inaccurate prefetching can undermine energy savings by incurring unnecessary remote transfers and cache pollution, increasing both bandwidth and memory energy consumption without improving performance~\cite{almaruf2020leap,atc23_cxlssd}.
Existing far-memory prefetching systems only perform simple rule-based predictions, such as sequential and stride, to achieve the accuracy and efficiency requirements~\cite{fastswap,almaruf2020leap,infiniswap}.
In real-world applications, complex access patterns such as pointer chasing and tree traversals amplify the cost of on-demand far-memory fetches, causing significant slowdowns in many data-center workloads~\cite{asplos25pulse,cheng2024characterizingdilemmaperformanceindex}.


We argue that, to harvest the energy benefit with far memory,
prefetching demands both \textit{prediction accuracy} and \textit{execution efficiency}.
We propose {\em \sys}, a Linux-based far-memory system encorporating a novel deep-learning (DL) guided prefetching mechanism.
Our key insight is that memory access behavior is governed by both application semantics (\eg, algorithmic logic)
and input-dependent runtime context (\eg, memory layout).
The semantics tend to generalize across inputs and can be learned offline, but the actual memory addresses are input-specific and best handled at runtime.
\sys\ exploits this separation by training a DL model to learn semantic patterns offline while tracking and predicting the actual access address efficiently at runtime.

We implement \sys’s prediction and prefetching system in the Linux kernel.
We build a predictor model based on RetNet~\cite{retnet}, a transformer~\cite{vaswani2017attention} variant with constant-time inference. Our implementation achieves an average inference latency of 1\mus\ on a single CPU core. 
\sys{} leverages an energy- and carbon-efficient setup by reusing older-generation devices as far memory, and outperforms state-of-the-art systems by up to 42\%.

{
\begin{figure*}[th]
\begin{minipage}{0.48\textwidth}
\begin{center}
\vspace{0.15in}
\centerline{\includegraphics[width=0.98\columnwidth]{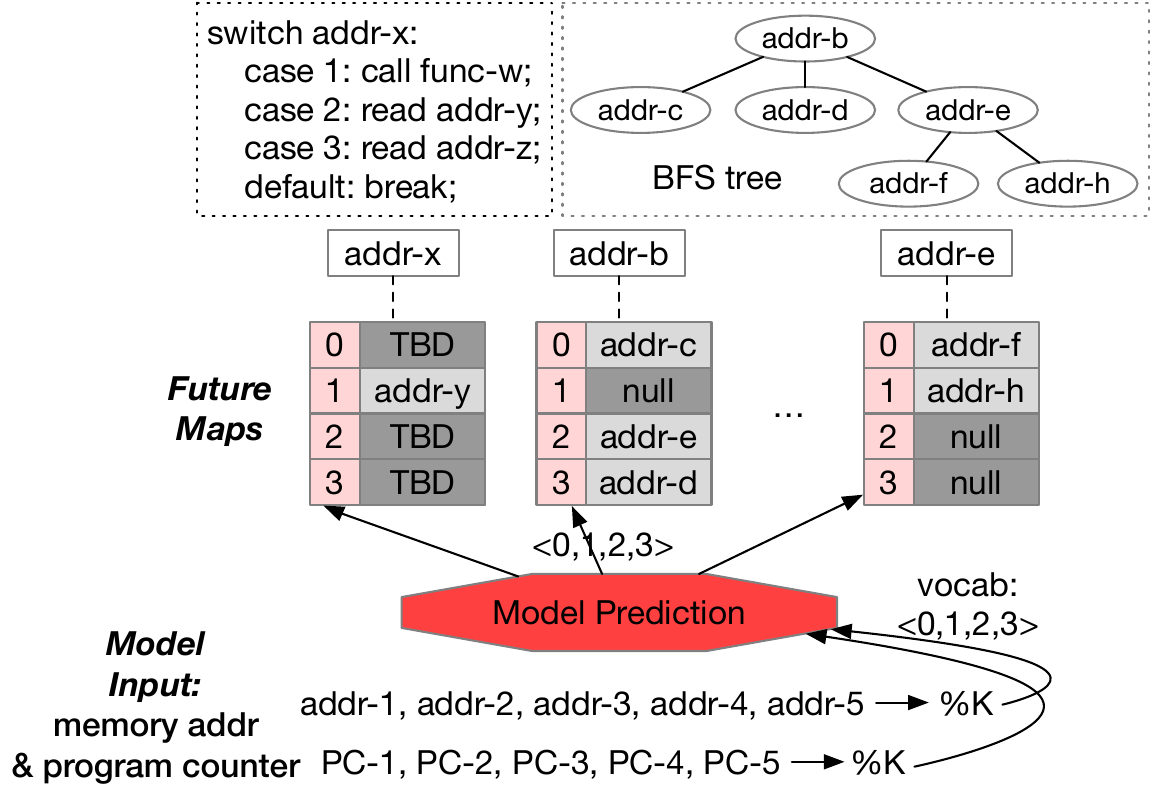}}
\mycaption{fig-futuremap}{\sys\ prediction representation}
{
An example of vocabulary size ($K$) being 4. The top part shows code/algorithm corresponding to the accesses of chunks \texttt{addr-x}, \texttt{addr-b}, and \texttt{addr-e}. The bottom shows the input to the model: the chunk addresses and PCs of the 5 previous misses.
}
\end{center}
\end{minipage}
\hfill
\vspace{-0.1in}
\begin{minipage}{0.48\textwidth}
\begin{center}
\vspace{0.15in}
\centerline{\includegraphics[width=\columnwidth]{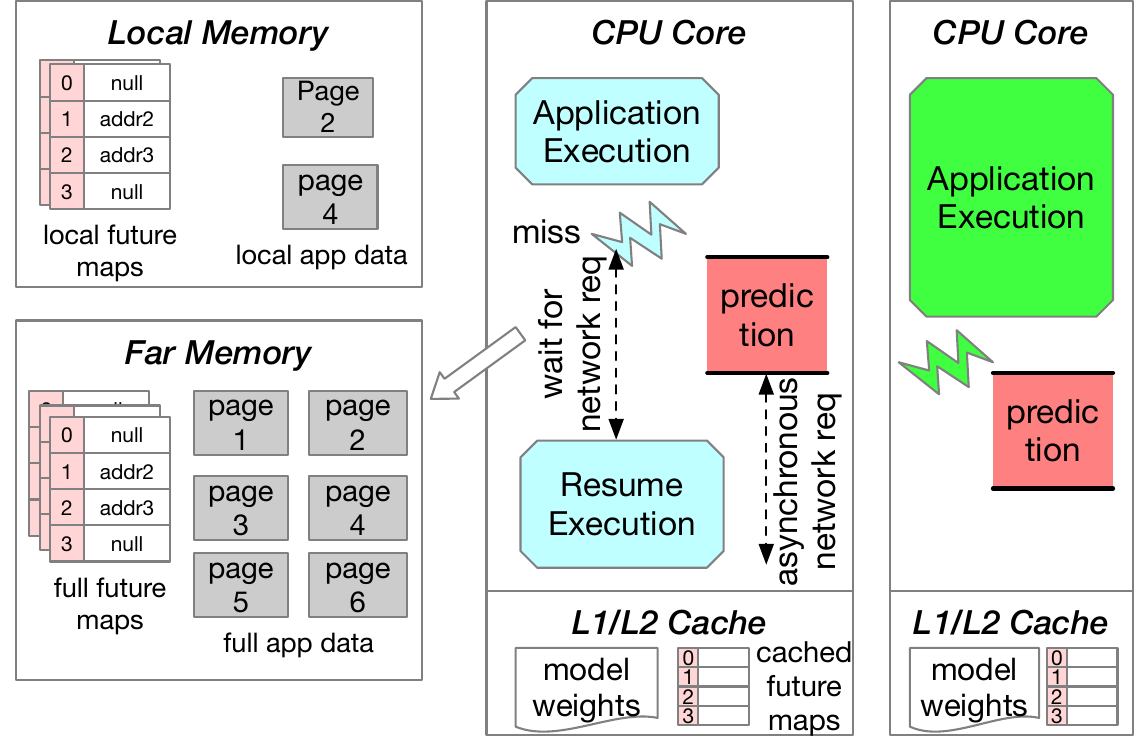}}
\vspace{0.04in}
\mycaption{fig-threadmodel}{\sys\ threading model}
{
The left panel shows the \sys\ system’s memory layout. The middle panel shows local memory misses during single-core execution, where \sys\ waits only for on-demand requests. The right panel shows that prefetch hits incur no network request.
}
\end{center}
\end{minipage}
\end{figure*}
}

\section{Two Key Ideas}



\paragraph{Decoupling Access Pattern and Addresses Prediction}
The first idea centers on improving prediction through a deeper understanding of the memory access process.
We observe that an application's memory access patterns are often repetitive due to code patterns like loops and recursion.
However, these patterns are typically complex and nonlinear~\cite{pmlr-v80-hashemi18a}, making them difficult to model with simple rules or heuristics.
DL models, on the other hand, are well-suited to capturing long-range, nonlinear dependencies.
To maintain efficiency, specially with smaller models that can run within a single CPU core and are trained offline, we must simplify the problem space and minimize runtime variability.

Our idea is to decouple application memory access semantics from the actual runtime memory layout by using DL prediction for the former and mapping tables for the latter. Specifically, we use a small DL model to predict memory access relationships in terms of {\em abstract ordinals}—--representing the possible memory-access outcomes after a short history of memory behavior—--rather than concrete memory addresses or offsets, which are highly input- and environment-dependent.
At runtime, we construct {\em future maps}: mapping tables that resolve these predicted ordinals to actual memory addresses observed during the program’s first access, thus capturing the true memory layout dynamically. Figure~\ref{fig-futuremap} illustrates this idea with an example.

\paragraph{Overlapping I/O and  In-Kernel Model Inference}
Our second main idea focuses on the execution efficient of the DL-based prediction framework.
The latency introduced by PCIe transfers and kernel launch overhead makes GPU-based prediction unsuitable for our use case~\cite{zhang2019cudaoverhead}. Thus, \sys\ performs its model prediction in CPU. To avoid the energy and performance cost of additional CPU cores, \sys\ performs prediction for an application on the same core it runs on.
Our idea to avoid application performance overhead is to hide model prediction behind far-memory I/O time.
Specifically, \sys\ performs prediction when a foreground page fault is being handled with far-memory data read. We then issue asynchronous prefetch requests in the background.

\section{\sys\ Design}

\sys\ is a swap-based far-memory system that prefetches
memory pages (instead of individual words) from far memory with an
DL-based memory access predictor. \sys\ consists of
an offline training component and an online predictor and
prefetcher component sitting in the Linux kernel’s swap
system, as seen in Figure~\ref{fig-threadmodel}.

\paragraph{Model Inputs}
\sys\ uses page miss history as the input to the DL model instead of full memory access history.
The reason is that, by being in the swap system, \sys\ can observe and log miss addresses on every page fault without incurring additional overhead. In contrast, capturing the full memory access stream would introduce substantial runtime overhead and is therefore avoided.
In addition to using page miss addresses, we associate every miss with the faulting program counter (PC), as doing so can incorporate program execution information with memory access history, and recording and using PC incurs no additional overhead.

To fit the two types of inputs into the vocabulary, we take the mod of their value to the vocabulary size, $K$.
Just like graph traversal can be recorded using neighbor IDs instead of absolute node IDs, page accesses can be tracked using page mod $K$ to capture relative page transition.
We then use a history sequence of $h$ pairs of the modulo of miss page address and PCs as the model input, as shown at the bottom of Figure~\ref{fig-futuremap}.
Even though taking a mod is inevitably a lossy process, a history sequence and two types of information still allow our model to make accurate predictions.

\paragraph{Model Outputs and Future Maps}
We choose to predict page misses (\ie, accesses to memory pages not in local memory), rather than attempting to predict every individual memory access--—which would be computationally intensive and unnecessary.
Essentially, \sys\ uses page miss sequence in recent history to predict page miss sequence in the future.
This approach significantly reduces the computational load on the DL model and the monitoring overhead.

Our solution is to label possible outcomes of memory access as ordinals. 
Specifically, we record a vocabulary size (\ie, $K$) of possible next memory page misses after a miss happens at page $X$. 
Based on the model inputs as described above, our model predicts an ordinal from 0 to $K-1$, corresponding to one of the likely next page misses.
We dynamically maintain a {\em future map} for each page $X$.
Each entry in the future map represents one possible page to be accessed after the miss of page $X$, and the value of the entry is the runtime virtual memory address of the page.
A null future map entry represents an outcome that has not occured at the runtime yet.

\paragraph{Vocabulary Size}
Naturally, a program can have fewer or more possible memory pages to access than $K$ after a page is accessed.
If there are fewer possibilities (\eg, Figure~\ref{fig-futuremap}, pages \texttt{addr-b} only have three possible outcomes), the model will just not yield the remaining as a predicted value.
If there are more possibilities than $K$, the model will not properly capture the less frequently occurring accesses.
We set the default configurable $K$ to 64, which balances memory overhead with prediction accuracy.
This choice is effective because small future maps can fit into CPU L1/L2 caches, reducing prediction latency. 
In addition, applications with repeatable behavior typically exhibit only a few possible outcomes after a page fault~\cite{duong2024twilight}.
Finally, since memory allocators often place nearby requests within the same memory page, $K=64$ is sufficient to capture access patterns across pages.

\section{Evaluation Results}

\paragraph{Implementation}
We integrated \sys{} into the Linux kernel and implemented RetNet in kernel space with AVX-512 instructions on x86.
The model consists of two layers, each with a hidden dimension of 8, resulting in 2.5K trainable parameters. This compact size allows the parameters to reside entirely in CPU cache, enabling an average inference latency of 1\mus. 

\paragraph{Experiment Setup}
We evaluate \sys\ on our private clusters,
where a compute node, equipped with a 28-core Intel Xeon Gold 5512U CPU and DDR5 memory, is connected via 100Gbps RDMA to a memory node as far memory. 
The memory node uses older generation of CPU and memory, a 16-core Intel Xeon Gold 5218 CPU with DDR4 memory. This emulates a common data center scenario where new servers leverage older servers’ memory for additional capacity, while being carbon-efficient.

We compare \sys\ with two baselines: FastSwap~\cite{fastswap} and Hermit~\cite{hermit}.
FastSwap is a swap-based far-memory system implemented in the Linux kernel.
Hermit builds on top of FastSwap and improves its swap-out procedure to avoid swap-out being the application performance bottleneck.
Both systems use the Linux prefetching policy, which only follows simple and strict rules for issuing prefetches for sequential accesses.
We evaluate \sys\ on three applications: XGBoost~\cite{chen2016xgboost}, a machine learning framework for gradient boosted decision trees (GBDTs), PageRank in GAP benchmark suite~\cite{beamer2017gapbenchmarksuite}, MCF in SPEC 2006 benchmark~\cite{spec2006_429mcf}.
These three applications serve to illustrate a representative subset of data analytics and machine learning applications, highlighting the types of computational patterns and data access behaviors that \sys\ is designed to handle efficiently.

{
\begin{figure*}[th]
\begin{minipage}{0.32\textwidth}
\begin{center}
\centerline{\includegraphics[width=\columnwidth]{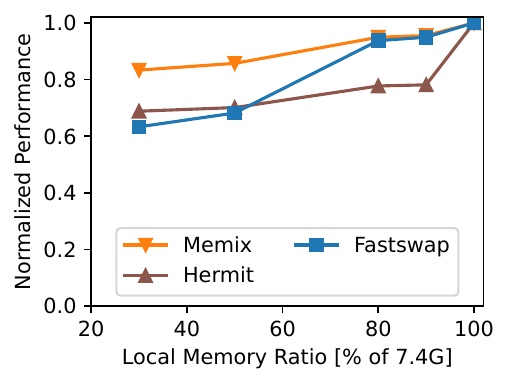}}
\mycaption{fig-overall-xgboost}{\small XGBoost performance}
{
}
\end{center}
\end{minipage}
\hspace{0.01\textwidth}
\vspace{-0.1in}
\begin{minipage}{0.32\textwidth}
\begin{center}
\centerline{\includegraphics[width=\columnwidth]{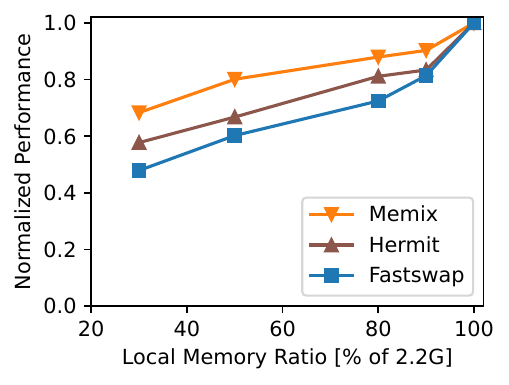}}
\mycaption{fig-overall-pr}{\small Pagerank performance}
{
}
\end{center}
\end{minipage}
\hspace{0.01\textwidth}
\begin{minipage}{0.32\textwidth}
\begin{center}
\centerline{\includegraphics[width=\columnwidth]{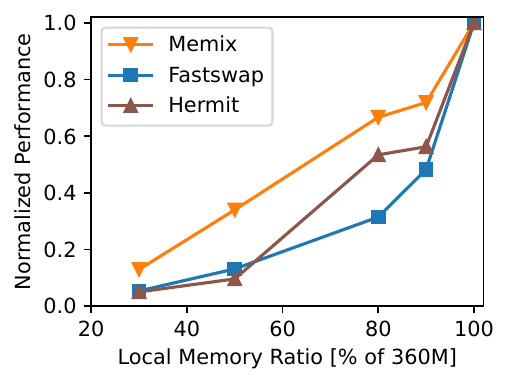}}
\mycaption{fig-overall-mcf}{\small MCF performance}
{
}
\end{center}
\end{minipage}
\vspace{0.1in}
\end{figure*}
}

\paragraph{Application performance}
Figures~\ref{fig-overall-xgboost}, ~\ref{fig-overall-pr}, ~\ref{fig-overall-mcf} present the end-to-end application performance of XGBoost, PageRank, and MCF compared to state-of-the-art systems.
These preliminary results show that \sys\ outperforms existing systems by up to 42\%.
For each set of experiments, we change the application server's local memory size from 30\% to 90\% of the total application memory size (X axis)
and measure the total application execution time (Y axis).
For each result, we normalize the application execution time against that of running at full local-memory capacity, and higher Y-axis values are better. 
Comparing to both systems, \sys\ generates up to 3x more useful prefetch in 30\% local memory size.

The improved performance of \sys\ yields energy savings.
By reducing memory stalls and enabling faster data access, CPUs spend more time idling and can enter low-power states more often.
In disaggregated memory pools, centralizing far memory reduces overprovisioning and eliminates duplicated data across servers, further lowering energy use. 
Consequently, the observed speedup enhances both throughput and energy efficiency, making \sys\ beneficial for performance and energy optimization in data centers.

\section{Discussion and Conclusion}
We presented \sys, a DL-based far-memory prefetching
system in the Linux kernel.
\sys’s core idea is to decouple the learning of application semantics from the runtime capturing of memory accesses.
By doing so, \sys\ achieves overall application performance benefits over two recent far-memory systems. 
Below we further discuss our setup, findings and future directions:

\paragraph{Target Applications}
Applications with repeated memory access paths are well-suited for \sys.
Most memory-intensive applications such as graph processing, data analytics, machine learning follow this.
Hash function is a counterexample, as the memory accessed is purely dependent on the hash function, not past sequences.

\paragraph{Target Environment}
\sys{} is targeting data center, where applications are used in enterprise only; typically ranging from tens to hundreds of applications.
We only need to train once per application or when its input drifts drastically.
For all the applications in the paper, our training time was around 30 minutes on a single A6000 GPU.

\paragraph{Lesson Learned}
\sys{}'s approach of separating access patterns and overlapping interference reveals further opportunities, one critical observation is \textit{swap-out management}. Our aggressive prefetching boosts performance but also increases swap-out activity, as each prefetch requires an empty page frame for new data. Even with perfect prefetching, evicting useful pages can degrade performance due to flushed data. This underscores that swap-out policy is equally critical and could be jointly optimized with I/O and prefetching, potentially guided by deep learning.

\paragraph{Future Directions}
CXL (Compute Express Link) \cite{CXL} enables memory expansion by attaching standalone memory devices over a PCIe-like interconnect, even with heterogeneous memory controllers (e.g., DDR4 via CXL with a DDR5 CPU). Similarly, CXL incurs higher latency than DRAM and thus sensitive to memory placement and prefetching. Its effectiveness relies on application semantic and placement-aware management, which \sys{} could be extended to support.

\section{Acknowledgment}
We would like to thank Ryan Lee, Geoff Voelker, Zijian He, Vikranth Srivatsa, and Reyna Abhyankar for their valuable contributions and feedback on this paper.
This material is based upon work supported by funding from PRISM center (part of SRC's JUMP 2.0) and gifts from AWS, Google, and Meta. Any opinions, findings, conclusions, or recommendations expressed in this material are those of the authors and do not necessarily reflect the views of these institutions.

\clearpage
\begin{small}
  \bibliographystyle{unsrt}
  \bibliography{paper}
\end{small}










\end{document}